\documentclass[10pt]{iopart}
\usepackage[T1]{fontenc}
\usepackage[utf8]{inputenc}
\usepackage{babel}
\usepackage{verbatim}
\usepackage{float}
\usepackage{amsmath}
\usepackage{amssymb}
\usepackage{graphicx}
\usepackage{booktabs}
\usepackage{graphicx}
\usepackage{subscript}
\usepackage[unicode=true,pdfusetitle,
bookmarks=true,bookmarksnumbered=false,bookmarksopen=false,
breaklinks=false,pdfborder={0 0 1},backref=false,colorlinks=false]
{hyperref}
\usepackage{xcolor}

\makeatletter

\newcommand{\lyxmathsym}[1]{\ifmmode\begingroup\def\b@ld{bold}
	\text{\ifx\math@version\b@ld\bfseries\fi#1}\endgroup\else#1\fi}

\usepackage{geometry} 
\usepackage{amsmath}
\usepackage{xparse}

\newmuskip\pFqmuskip
\newcommand*\pFq[6][8]{%
	\begingroup 
	\pFqmuskip=#1mu\relax
	\mathchardef\normalcomma=\mathcode`,
	\mathcode`\,=\string"8000
	\begingroup\lccode`\~=`\,
	\lowercase{\endgroup\let~}\pFqcomma
	{}_{#2}F_{#3}{\left[\genfrac..{0pt}{}{#4}{#5};#6\right]}%
	\endgroup
}
\newcommand{\pFqcomma}{{\normalcomma}\mskip\pFqmuskip}

\ExplSyntaxOn
\NewDocumentCommand{\MeijerG}{smmmm}
{
	\IfBooleanTF{#1}
	{
		\vic_meijerg:nnnnnn { #2 } { #3 } { #4 } { #5 } { small } { }
	}
	{
		\vic_meijerg:nnnnnn { #2 } { #3 } { #4 } { #5 } { } { \; }
	}
}

\seq_new:N \l__vic_meijerg_args_in_seq
\seq_new:N \l__vic_meijerg_args_out_seq

\cs_new_protected:Nn \vic_meijerg:nnnnnn
{
	\seq_set_split:Nnn \l__vic_meijerg_args_in_seq { | } { #3 }
	\seq_clear:N \l__vic_meijerg_args_out_seq  
	\seq_map_inline:Nn \l__vic_meijerg_args_in_seq
	{
		\seq_put_right:Nn \l__vic_meijerg_args_out_seq
		{
			\begin{#5matrix} ##1 \end{#5matrix}
		}
	}
	G\sp{#1}\sb{#2}
	\left(
	\seq_use:Nn \l__vic_meijerg_args_out_seq { #6\middle|#6 }
	#6\middle|#6
	#4
	\right)
}
\ExplSyntaxOff

\usepackage{babel}

\begin{document}
\title{Excitonic instability in transition metal dichalcogenides}
\author{M. F. C. Martins Quintela$^{1,2}$, A. T. Costa$^{2}$, N. M. R. Peres$^{1,2}$}
\address{$^{1}$Department of Physics and Centre of Physics of the Universites of Minho and Porto (CF--UM--UP), Campus of Gualtar, 4710-057, Braga, Portugal}
\address{$^{2}$International Iberian Nanotechnology Laboratory (INL), Av. Mestre Jos{\'e} Veiga, 4715-330, Braga, Portugal}
\ead{mfcmquintela@gmail.com}
\begin{abstract}
When transition-metal dichalcogenide monolayers lack inversion
symmetry, their low-energy single particle spectrum can described 
by tilted massive Dirac Hamiltonians. The so-called Janus materials
fall into that category. Inversion symmetry can also be broken by 
the application of out-of-plane electric fields, or by the mere 
presence of a substrate. Here we explore the properties of excitons in TMDC 
monolayers lacking inversion symmetry. We find that exciton binding 
energies can be larger than the electronic band gap, making such 
materials promising candidates to host the elusive exciton insulator 
phase. We also investigate the excitonic contribution to their
optical conductivity and discuss the associated optical selection 
rules.

\end{abstract}

\submitto{\JPCM}
\maketitle
\ioptwocol

\section{Introduction}

Janus transition metal dichalcogenide (TMD) monolayers are a new type of two--dimensional materials,  recently synthesized  \cite{Lu2017,Zhang2017} in the form of $ \mathrm{MoSSe} $. In this material, one atomic layer of $ \mathrm{Mo} $ is encapsulated by two different chalcogen layers, namely $ \mathrm{S} $ and $ \mathrm{Se} $. This creates an asymmetry in the direction
perpendicular to the plane of the structure, resulting in a dipole moment pointing from the $ \mathrm{Se} $ to the $ \mathrm{S} $ layer \cite{Anders2020_PhD,doi:10.1021/acs.nanolett.0c03412}. 
These first two independent experiments \cite{Lu2017,Zhang2017} triggered a large wave of theoretical studies\cite{Tang2022} into the electronic band structure\cite{Zhang2017}, anisotropic elasticity and transport\cite{Zhang2022}, as well as piezoelectric \cite{Dong2017}, pyroelectric \cite{PhysRevLett.120.207602}, spin \cite{PhysRevB.97.235404}, and photocarrier \cite{C8NR04568B}  properties of Janus structures. Additionally, several potential applications have been studied, such as in water splitting \cite{C7TA10015A}, gas sensing \cite{C8TA08407F},  and photovoltaics \cite{C9CP02648G}. 

The band structure and the Berry curvature dipole~\cite{Berry1984,RevModPhys.82.1959,PhysRevLett.115.216806} of Janus TMD monolayers are well described by a 
tilted massive Dirac model~\cite{Joseph_2021,PhysRevLett.121.266601,Haastrup2018}. 
For instance, this two-band effective model captures the pronounced peak in the dipole component 
near the Fermi level of $ \mathrm{T'-WSTe} $, seen in DFT calculations~\cite{Joseph_2021}.
This feature is a consequence of a reduced band gap, which leads to a 
large Berry curvature~\cite{Joseph_2021}. Tilted Dirac cones are also seen
in $ 8-Pmmm $ borophene \cite{PhysRevB.94.165403,PhysRevB.96.155418,doi:10.1063/5.0007576}, 
where the low-energy regime near the Dirac points can be accurately described via an 
effective anisotropic tilted Hamiltonian \cite{PhysRevB.94.165403,PhysRevB.96.155418,PhysRevB.99.035415,PhysRevB.100.195420}.

Several properties of systems with tilted massive Dirac Hamiltonians have been studied, 
such as anomalous spin transport\cite{10.7566/JPSJ.91.023708}, topological 
properties\cite{PhysRevX.6.041069}, photoinduced anomalous~\cite{LI2020114092} 
and nonlinear Hall effects~\cite{PhysRevLett.121.266601}, 
quantum criticality~\cite{PhysRevResearch.2.013069}, 
chiral excitonic instabilities~\cite{PhysRevResearch.2.033479}, 
and orbital-selective photoexcitation \cite{Guan2021}. 
Additionally, the importance of spin-orbit coupling in layered organic salts 
has also been studied via a Hamiltonian with tilted band 
structure~\cite{PhysRevB.95.060404, 10.7566/JPSJ.87.075002}. 
Explicit inversion symmetry breaking and anisotropic terms are also 
present in other families of materials, such as Weyl 
semimetals~\cite{10.1126/sciadv.1501524} with broken tilt inversion symmetry, 
where second harmonic generation has been recently studied~\cite{Gao:21},
and type-II semi-Dirac semimetals~\cite{PhysRevLett.102.166803,PhysRevB.92.161115,PhysRevB.94.081103}, 
where it has been shown that linear and nonlinear anomalous Hall effects can be
manipulated via circularly polarized light \cite{PhysRevB.105.085124}. 

A defining feature of massive tilted Dirac cones is the tunability of
their electronic band gap, illustrated in Fig. (\ref{fig:band_structure}).
By tuning a single parameter, it is possible to go from a direct gap semiconductor
to a metal, passing through an indirect gap semiconductor and a semi-metal. 
\textit{Ab initio} calculations~\cite{Zhang2018} suggest that such tunability can be
achieved in real materials by the application of an out-of-plane electric field.
Thus, materials that host massive tilted Dirac cones are especially attractive as 
platforms on which new electronic phases can be found. One such phase is the
excitonic insulator, first predicted qualitatively by Mott in 1961 \cite{Mott1961}.  
The concept has been subsequently refined and put on firmer grounds by several
authors \cite{desCloizeaux1964,Kohn1967,Jerome1967,Halperin1968}.
In brief, when the binding energy of excitons surpasses the
value of the electronic band gap, the system becomes unstable against a 
proliferation of excitons\cite{Knox1963}. The true ground state of such a system
is an ``exciton condensate'', akin in many ways to the Cooper pair condensate
found in superconductors\cite{Jerome1967}. This analysis apply equally
to small gap semiconductors and semimetals, provided the overlap between
conduction and valence bands is small, such that screening of the Coulomb
interaction between electrons and holes is negligible. Although predicted
more than 60 years ago, the excitonic insulator has eluded conclusive
experimental observation until very recently. One of the main difficulties
is that the excitonic instability is frequently accompanied by structural
instabilities, associated with the softening of phonon modes, which are hard 
to disentangle from the former.

Here we look into the properties of excitons in 2D massive tilted
Dirac electrons. We show that the band gap can be tuned while
the exciton binding energy remains constant, moving the system
towards an excitonic instability. Importantly, as the magnitude of
the gap decreases, the maximum of the valence band and the minimum of
the conduction band shift in opposite dirrections in reciprocal space,
making the gap indirect. This guarantees that the dielectric function
of the material remains finite even as the gap approaches zero,
and the exciton binding energies are not strongly affected by the smallness
of the gap\cite{desCloizeaux1964}. It is also noteworthy that the excitonic
contribution to the conductivity is insensitive to the tilting parameter; thus,
at least in principle, one can expect a sudden change in transport properties
as the excitonic instability is reached by tuning of the tilting parameter.

This paper is structured as follows. In Sec. \ref{sec:2}, we discuss the 
tilted Dirac Hamiltonian \cite{PhysRevLett.121.266601}, illustrating the
features induced by the tilt parameter on the single particle electronic 
eigenstates and band structure. In Sec. \ref{sec:3}, we introduce the 
Bethe--Salpeter equation, briefly discussing the electrostatic potential 
coupling different bands. The screening length of the material is also 
introduced, and the influence of the tilt parameter on the excitonic states 
is analyzed. Finally, the optical selection rules and oscillator strengths 
are discussed, and the excitonic optical conductivity is computed. 

\section{Tilted Dirac Hamiltonian\label{sec:2}}

The effective two-band tilted Dirac Hamiltonian has been shown to capture the
essential features of both the low-energy band structure and the Berry curvature 
dipole moment of Janus TMD monolayers, such as $\mathrm{WSTe}$~\cite{Joseph_2021}.
This Hamiltonian includes an anisotropic term which preserves time-reversal symmetry 
but explicitly breaks inversion symmetry, tilting the band structure in a specific 
direction, here considered to be the $ x $-axis. The Hamiltonian can therefore be 
written as  \cite{PhysRevLett.115.216806,Joseph_2021,PhysRevLett.121.266601}
\begin{equation}
	\hat{\mathcal{H}}_{d}=tk^{x}\sigma_{0}+v\left(k^{y}\sigma_{x}+
        \eta k^{x}\sigma_{y}\right)+\left(m/2-\alpha k^{2}\right)\sigma_{z},\label{eq:tilted_ham}
\end{equation}
where $\left(k_{x},k_{y}\right)$ are the wave vectors, 
$k^{2}=k_{x}^{2}+k_{y}^{2}$, $\left(\sigma_{x},\sigma_{y},\sigma_{z}\right)$ 
are the Pauli matrices, $\sigma_{0}$ is the $2\times2$ identity matrix, 
$\eta=\pm1$ is a valley--like index, $m$ is the gap, and $t$ tilts the Hamiltonian in the $x$ direction. 
In Eq. (\ref{eq:tilted_ham}), the $\alpha$ parameter is introduced to regulate 
topological properties as $k\rightarrow\infty$ \cite{Shen2012}. For clarity, 
we also write this Hamiltonian in matrix form, where it is given by 
\begin{equation}
\hat{\mathcal{H}}_{d}=\left[\begin{array}{cc}
	tk_{x}+\left(\frac{m}{2}-\alpha k^{2}\right) & -iv\left(\eta k_{x}+ik_{y}\right)\\
	iv\left(\eta k_{x}-ik_{y}\right) & tk_{x}-\left(\frac{m}{2}-\alpha k^{2}\right)
\end{array}\right].
\end{equation}

\begin{figure*}
	\centering
	\includegraphics[scale=0.75]{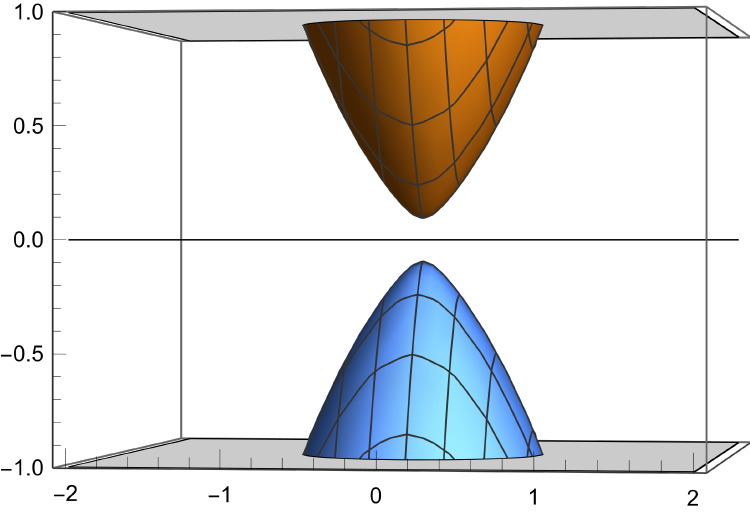}
	\quad\includegraphics[scale=0.75]{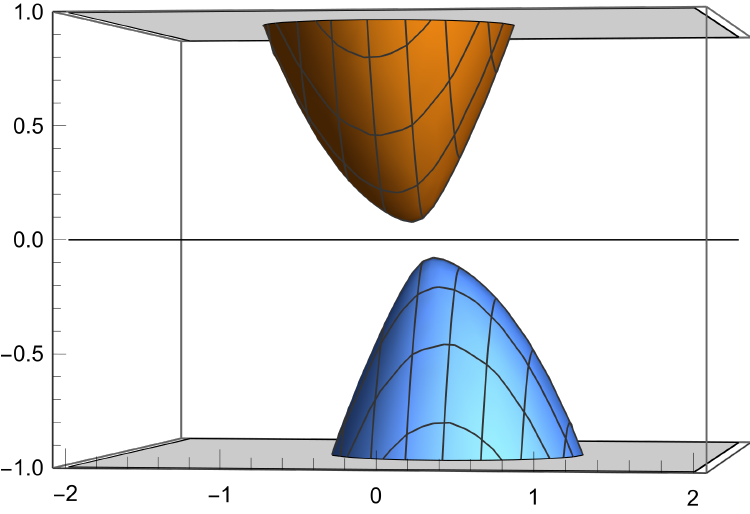}
	
	\includegraphics[scale=0.75]{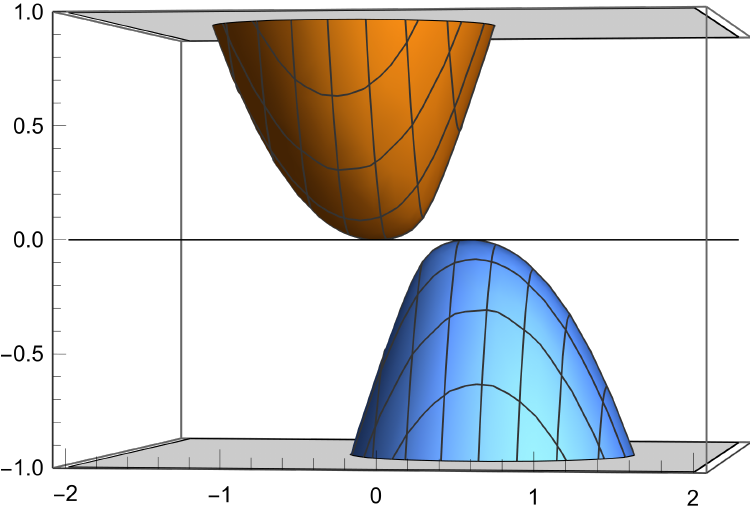}
	\quad\includegraphics[scale=0.75]{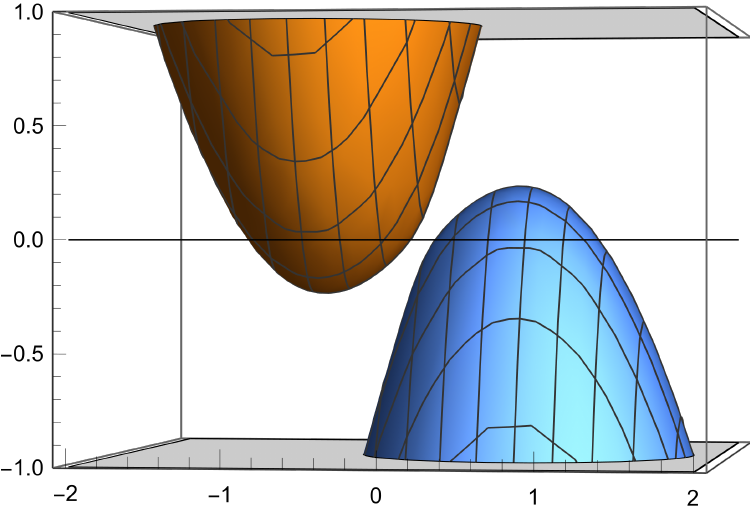}
	
	\caption{3D plot of the band structure of the tilted Dirac Hamiltonian of Eq. (\ref{eq:tilted_ham}) for $ t/v=0 $ (top-left), $ t/v=0.5 $ (top-right), $ t/v=1 $ (bottom-left), and $ t/v=1.5 $ (bottom-right). Labeled axes are the $ k_x $ axis (horizontal, in $ \text{\AA}^{-1} $) and energy (vertical, in $ \mathrm{eV} $).\label{fig:band_structure}}
\end{figure*}

The band dispersion of this model is given by 
\begin{equation}
	E_{\lambda}\left(k\right)=tk_{x}+\lambda\sqrt{k^{2}v^{2}+\left(m/2-\alpha k^{2}\right)^{2}},\label{eq:band_structr}
\end{equation}
with $\lambda=\pm1$ the conduction/valence band index. Choosing $\eta=-1$ as in \cite{Joseph_2021,PhysRevLett.121.266601}, the (non-normalized) eigenvectors are given by 

\begin{align}
	\left|u_{+}\left(k,\theta\right)\right\rangle &=\left[\begin{array}{c}
		i\frac{m-2\alpha k^{2}+\lambda\sqrt{k^{2}v^{2}+\left(m/2-\alpha k^{2}\right)^{2}}}{2kv}\\
		e^{i\theta}
	\end{array}\right], \nonumber\\ \left|u_{-}\left(k,\theta\right)\right\rangle &=\left[\begin{array}{c}
	ie^{-i\theta}\frac{m-2\alpha k^{2}+\lambda\sqrt{k^{2}v^{2}+\left(m/2-\alpha k^{2}\right)^{2}}}{2kv}\\
	1
\end{array}\right],\label{eq:eigenvectors}
\end{align}
where $ \theta=\arctan\left(k_y / k_x\right) $. By inspection of both Eq. (\ref{eq:band_structr}) 
and Eq.~(\ref{eq:eigenvectors}) it is immediately clear that the tilt parameter $ t $ influences 
neither the difference between the two bands (\emph{i.e.,} $ E_+\left(k\right) - E_-\left(k\right) $) 
nor the eigenstates of the Hamiltonian. Its only effect is to produce a tilt of the dispersion 
relation along the $ x $-axis. In Fig.~\ref{fig:band_structure} we show the dispersion relation 
for a few representative values of $t$. The other parameters have been chosen as 
$v=1\,\mathrm{eV}\text{\AA}$, $\alpha=1\,\mathrm{eV}\text{\AA}^{2}$ and 
$m=0.2\,\mathrm{eV}$\cite{PhysRevLett.121.266601}. For $t=0$ (top left) there is a gap at $k=0$, which
is continuously suppressed as $t\rightarrow v$. Notice also that, for any $ 0 < t < v$, the
gap becomes indirect. For $t/v \ge 1$ (bottom panels) the model describes a semimetal with 
zero ($t=v$) or small $t/v \gtrsim 1$ carrier density.

\begin{figure}
	\centering
	\includegraphics[scale=0.75]{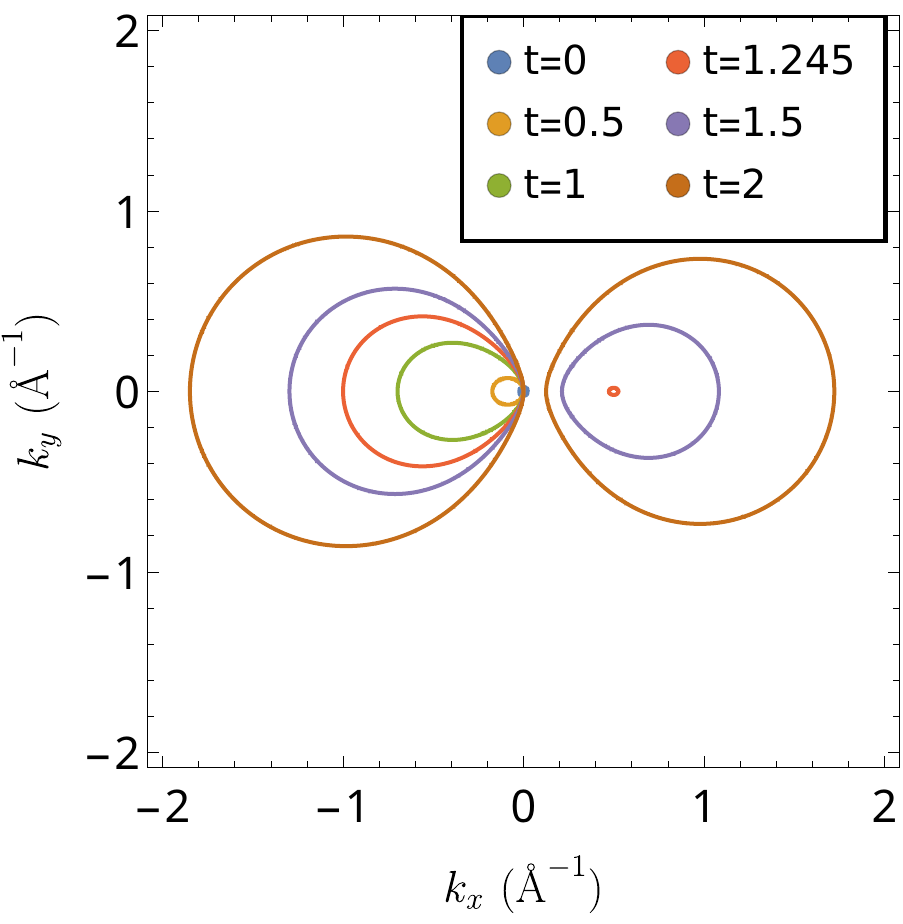}
	
	\caption{Fermi rings of the Hamiltonian of Eq. (\ref{eq:tilted_ham}) at $ E_F= m/2 $ for various values of $ t/v $. The rings associated with the conduction band are present on the left of the figure, while those associated with the valence bands are present on the right.\label{fig:fermi_surface}}
	
\end{figure}

In Fig. (\ref{fig:fermi_surface}), we plot the Fermi rings at a Fermi energy of $ E_F = m/2 $ 
for various values of $ t/v $. The top of the valence band crosses the $ E= \dfrac{m}{2} $ plane 
at roughly $ t/v\approx1.245 $, as can be seen from the sudden appearance of the Fermi ring regarding 
this band. The anisotropy introduced by the tilt of the bands along the $ k_x $ axis is also clear 
when compared with the full symmetry along the $ k_y $ axis.

\section{Excitons in tilted Dirac materials\label{sec:3}}

We now look at the excitons of the massive tilted Dirac fermion model.
Our approach is to solve the Bethe-Salpeter equation (BSE).
As neither the difference between the two bands nor the eigenvectors 
depend on $t$, we do not expect this parameter to play a part in the 
excitonic properties of the system. 

\subsection{Solving the Bethe-Salpeter equation}

The Bethe-Salpeter equation can be written in momentum 
space as \cite{PhysRevB.99.235433,PhysRevB.92.235432,PhysRevLett.120.087402,PhysRevB.104.115120}
\begin{align}
	&E \, \psi_{c;v}\left(\mathbf{k}\right)  =\left(E_{\mathbf{k}}^{c}-E_{\mathbf{k}}^{v}\right)\psi_{c;v}\left(\mathbf{k}\right)+\nonumber\\
	&\quad+\sum_{\mathbf{q}}V\left(\mathbf{k}-\mathbf{q}\right)\left\langle u_{\mathbf{k}}^{c}\mid u_{\mathbf{q}}^{c}\right\rangle \left\langle u_{\mathbf{q}}^{v}\mid u_{\mathbf{k}}^{v}\right\rangle\psi_{c;v}\left(\mathbf{q}\right), \label{eq:BSE-simplified}
\end{align}
where $ \psi_{c;v}\left(\mathbf{k}\right) $ is the excitonic wave function that we wish to 
obtain,  $\left|u_{\mathbf{k}}^{v/c}\right\rangle $ and $E_{\mathbf{k}}^{v/c}$ are the single 
particle electronic wave functions and energies, respectively, and $V\left(\mathbf{k}\right)$ 
is an electrostatic potential coupling different bands and thus capturing many-body effects 
including the intrinsic many-body nature of excitons. 

Assuming the excitons have a well-defined angular momentum $ \ell $, we can write their wave 
functions as $\psi_{c;v}\left(\mathbf{k}\right)=f_{c;v}\left(k\right)e^{i\ell\theta}$. Under 
this assumption, we can write the BSE as 
\begin{align}
	&E \, f_{c;v}\left(k\right)e^{i\ell\theta_k}  =\left(E_{\mathbf{k}}^{c}-E_{\mathbf{k}}^{v}\right) f_{c;v}\left(k\right)e^{i\ell\theta_k}+\nonumber\\
	&\quad+\sum_{\mathbf{q}}V\left(\mathbf{k}-\mathbf{q}\right)\left\langle u_{\mathbf{k}}^{c}\mid u_{\mathbf{q}}^{c}\right\rangle \left\langle u_{\mathbf{q}}^{v}\mid u_{\mathbf{k}}^{v}\right\rangle f_{c;v}\left(q\right)e^{i\ell\theta_q}. \label{eq:BSE-momentum}
\end{align}
Taking the thermodynamic limit and rearranging the complex exponentials, the Bethe-Salpeter 
equation is now given by
\begin{align}
	&E \, f_{c;v}\left(k\right)  =\left(E_{\mathbf{k}}^{c}-E_{\mathbf{k}}^{v}\right)f_{c;v}\left(k\right)-\int\frac{qdqd\theta_{q}}{4\pi^{2}}\left[V\left(\mathbf{k}-\mathbf{q}\right)\right.\nonumber\\
	&\qquad\left.*\left\langle u_{\mathbf{k}}^{c}\mid u_{\mathbf{q}}^{c}\right\rangle \left\langle u_{\mathbf{q}}^{v}\mid u_{\mathbf{k}}^{v}\right\rangle f_{c;v}\left(q\right)e^{i\ell\left(\theta_{q}-\theta_{k}\right)}\right].\label{eq:BSE-limit}
\end{align}
The process of solving the Bethe-Salpeter equation in monolayer and multilayer systems 
has been thoroughly discussed recently \cite{PhysRevB.105.045411,Quintela2022}, so we 
will not go into the details of the calculations. 

We consider the electrostatic interaction to be given by the Rytova-Keldysh 
potential \cite{rytova1967,keldysh1979coulomb}, obtained by solving the Poisson 
equation for a charge embedded in a thin film of vanishing thickness. In momentum 
space, this potential is given by 
\begin{equation}
	V\left(\mathbf{k}\right)=2\pi \frac{\hbar c\alpha}{\epsilon}\frac{1}{k\left(1+r_{0}k\right)},
\end{equation}
where $\alpha=1/137$ is the fine-structure constant and $\epsilon$ the mean dielectric 
constant of the medium above/below the monolayer, here considered to be either hexagonal 
boron-bitride (hBN) or quartz. The parameter $r_{0}$ corresponds to an in-plane screening length 
related to the 2D polarizability of the material. It can be calculated from the single 
particle Hamiltonian of the system\cite{PhysRevB.99.035429}, for $t<v$, as
\begin{align}
	r_{0} & =\frac{\hbar^{3}c\alpha}{\pi m_{0}^{2}}\int\frac{\left|\left\langle u_{\mathbf{k}}^{c}\left|P_{x}\right|u_{\mathbf{k}}^{v}\right\rangle \right|^{2}}{\left[E_{c}\left(k\right)-E_{v}\left(k\right)\right]^{3}}k\,dk\,d\theta,\label{eq:screen_2bands}
\end{align}
with $ m_{0} $ the free electron mass, although \emph{ab initio} calculations might be necessary for accurate computation 
of $ r_0 $ depending on the material\cite{acs.nanolett.9b02982}.

\subsection{Influence of the tilt parameter}
When computing the momentum matrix element present in Eq. (\ref{eq:screen_2bands}), a dependence on $t$ is only present on its $ x $ component as  
\begin{equation}
	P_{x}=\frac{m_{0}}{\hbar}\frac{\partial}{\partial k_{x}}\hat{\mathcal{H}}_{d}=\frac{m_{0}}{\hbar}\left[\begin{array}{cc}
		t-2k_{x}\alpha & -iv\eta\\
		iv\eta & t+2k_{x}\alpha
	\end{array}\right].\label{eq:momentum_matrix}
\end{equation}
However, the diagonal terms proportional to $t$ are canceled by the orthogonality relation of the two eigenvectors when $\left\langle u_{\mathbf{k}}^{c}\left|P_{x}\right|u_{\mathbf{k}}^{v}\right\rangle$ is computed. Explicitly, this term reads
\begin{equation}
	e^{-i\theta}t\left[a_{+}^{\dagger}\left(k\right)a_{-}\left(k\right)+b_{+}^{\dagger}\left(k\right)b_{-}\left(k\right)\right]=0,
\end{equation}
with $ a_{\pm} $ and $ b_{\pm} $ the normalized spinor components of the eigenvectors in Eq. (\ref{eq:eigenvectors}), written generically as 
\begin{align}
	\left|u_{+}\left(k,\theta\right)\right\rangle  & =\left[\begin{array}{c}
		a_{+}\left(k\right)\\
		b_{+}\left(k\right)e^{i\theta}
	\end{array}\right], \nonumber\\	\left|u_{-}\left(k,\theta\right)\right\rangle  & =\left[\begin{array}{c}
		a_{-}\left(k\right)e^{-i\theta}\\
		b_{-}\left(k\right)
	\end{array}\right].\label{eq:eigenvectors_simple}
\end{align}

This cancellation, together with the fact that neither the difference of energy between the two 
bands nor the eigenvectors themselves depend on $ t $, implies that the $ t $ parameter will not 
change the results obtained from solving the Bethe-Salpeter equation. As such, the obtained 
excitonic states will be independent of the tilt parameter $ t $. 

\begin{figure}
\includegraphics[width=0.5\columnwidth]{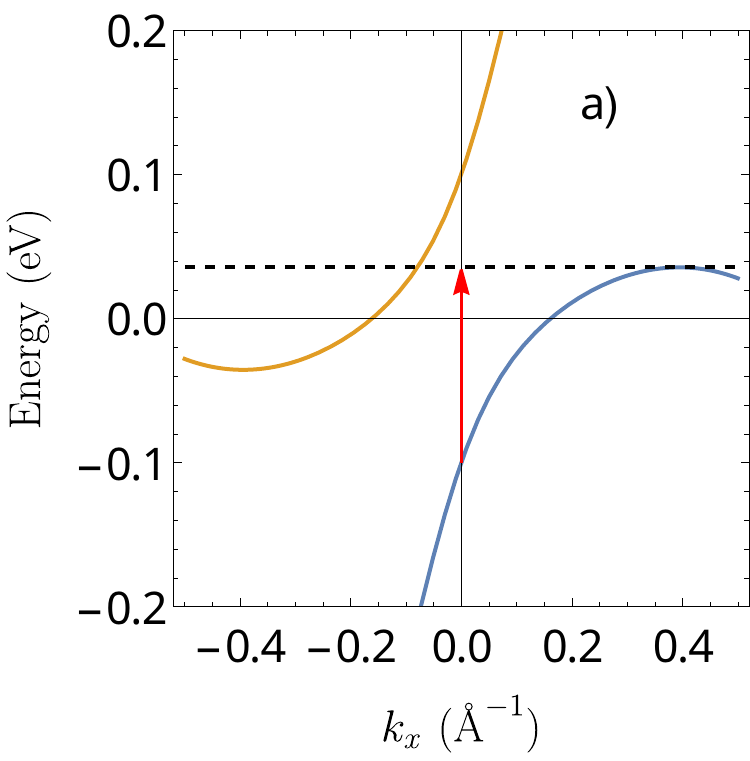}
\includegraphics[width=0.5\columnwidth]{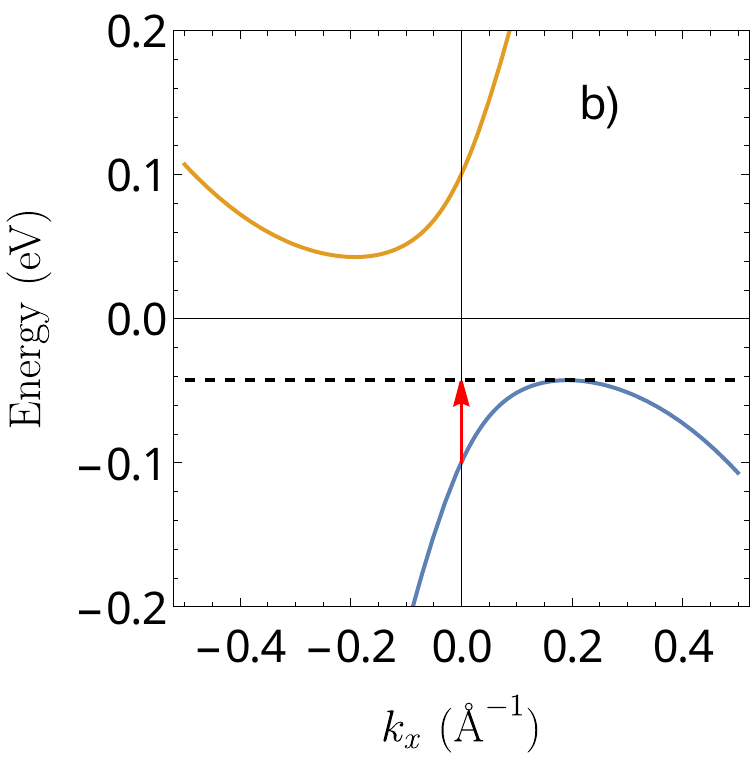}
\caption{Band structure of the tilted model for two different
values of the tilt parameter, $t=1.1v$ (a) and $t=0.83v$ (b).
In each case we show the excitonic level (dashed lines) for 
different substrates, hBN (a) and quartz (b). The values chosen 
for $t$ correspond to the onset of the excitonic instability for 
each case.}
\label{fig:instability}
\end{figure}

First considering the TMD encapsulated in hBN, the energies of first and second $ s $-series states are, respectively, 
$ E_{1s}=134\,\mathrm{meV} $ and $ E_{2s}=176\,\mathrm{meV} $. 
When $t\approx 1.1v$, the top of the valence band crosses the excitonic level,
as shown in Fig. (\ref{fig:instability}-a).
This marks the onset of the instability against the spontaneous formation of
excitons. Although the system is a semi-metal for this value of the
ratio $t/v$, the carrier density is still very small, indicating that
screening is still weak and the long-range character of the electron-hole
interaction should be preserved.

By changing the material by which the TMD is encapsulated, it is possible
to tune the exciton binding energy. For instance, by replacing hBN by quartz
(whose relative dielectric constant is 3.8 \cite{Serway2003}), the energy of 
the $1s$ exciton is $ E_{1s}= 56.6\,\mathrm{meV}$.
In this case, the onset of the excitonic
instability happens for $t\approx 0.83v$, well into the semiconducting regime,
as shown in Fig. (\ref{fig:instability}-b).
The fact that, for all $t\ne 0$, the gap is indirect, guarantees that
the renormalization of the exciton binding energy and the dielectric function
are small even for an arbitrarily small gap\cite{desCloizeaux1964}.

\begin{figure*}
	\qquad\qquad\includegraphics[scale=0.7]{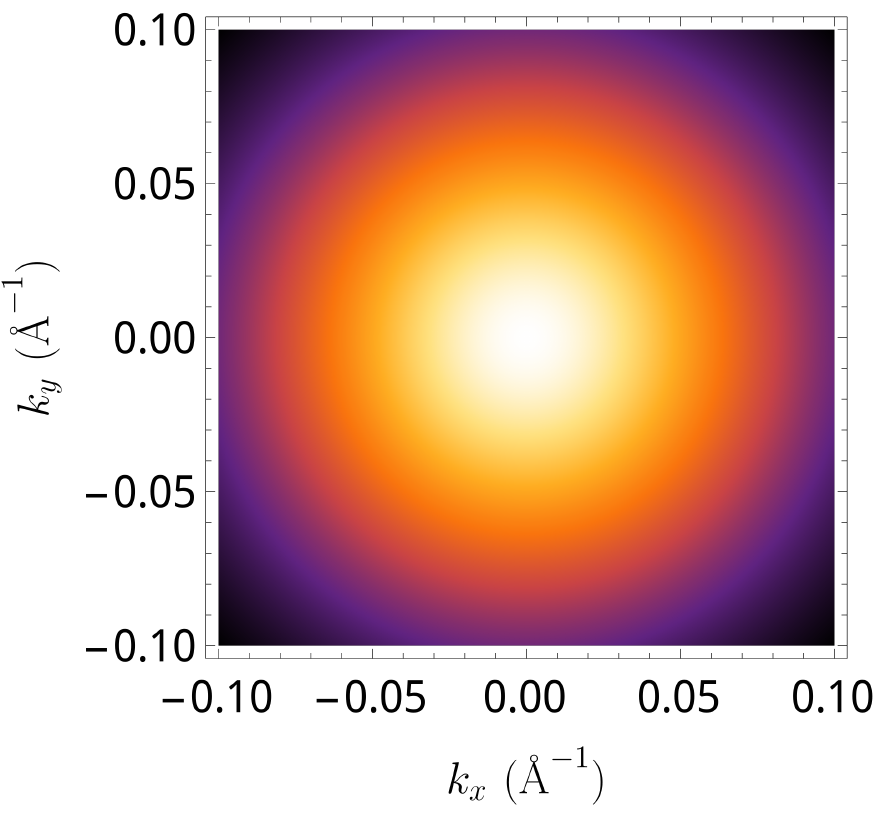}
	\quad\includegraphics[scale=0.7]{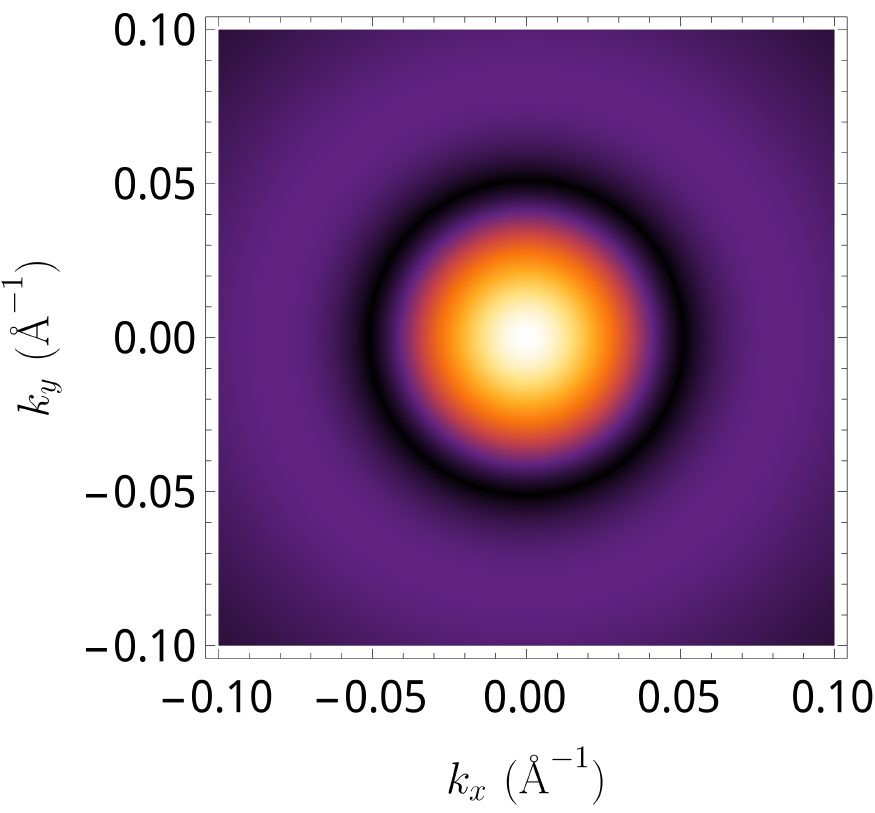}
	
	\caption{Absolute value squared of the wave functions of the two lowest energy excitonic $ s $-series states centered at $ k=0 $ considering the TMD encapsulated in quartz. \label{fig:excitonic_states}}
	
\end{figure*}

In Fig.~(\ref{fig:excitonic_states}) 
we plot the absolute value squared of the first two $ s $-series excitonic wave functions where the TMD has been encapsulated in quartz. These plots are centered at $ k=0 $ for a square region of side $ 20 \,\text{\AA}^{-1}$. 

\subsection{Excitonic Conductivity}

In the dipole approximation, and considering normal incidence, the optical conductivity is given by\cite{PhysRevB.92.235432} 
\begin{equation}
	\sigma_{\alpha,\beta}^{(1)}\left(\hbar\omega\right)\propto \sum_{n} E_{n}\frac{\boldsymbol{\Omega}_{n,\alpha} \boldsymbol{\Omega}_{n,\beta}^{*}}{E_{n}-\hbar \omega-i\Gamma_n}+(\omega \rightarrow-\omega)^{*},\label{eq:opt_cond}
\end{equation}
where the sum over $n$ represents the sum over excitonic states with energy $ E_n $ and wave function $ \psi_{n} $, and $ \Gamma_n $ is a phenomenological broadening parameter considered to be $ n $-dependent in a similar fashion as \cite{PhysRevB.105.045411}. In Eq. (\ref{eq:opt_cond}), $\boldsymbol{\Omega}_{n,\alpha}$ is defined as
\begin{equation}
	\boldsymbol{\Omega}_{n,\alpha}=\sum_{\mathbf{k}}\psi_{n}\left(\mathbf{k}\right)\left\langle u_{\mathbf{k}}^{v}\left|\mathbf{r}_{\alpha}\right|u_{\mathbf{k}}^{c}\right\rangle ,\label{eq:tg_pedersen_OMEGA}
\end{equation}
with $\left\langle u_{\mathbf{k}}^{v}\left|\mathbf{r}_{\alpha}\right|u_{\mathbf{k}}^{c}\right\rangle$ the interband dipole operator matrix element in the $ \alpha $ direction, obtained using the relation
\begin{equation}
	\left\langle u_{\mathbf{k}}^{v}\left|\mathbf{r}_{\alpha}\right|u_{\mathbf{k}}^{c}\right\rangle =\frac{\left\langle u_{\mathbf{k}}^{v}\left|\left[H,\mathbf{r}_{\alpha}\right]\right|u_{\mathbf{k}}^{c}\right\rangle }{E_{k}^{v}-E_{k}^{c}}.\label{eq:selection_rules_commutator}
\end{equation}
Inserting this relation into Eq. (\ref{eq:opt_cond}), we then write the excitonic $ xx $-conductivity as 
\begin{align}
	&\sigma_{xx}^{(1)}(\omega)=\frac{e^{2}}{4 \pi^{2} i \hbar}\sum_{n}\frac{E_{n}\left|\int\psi_{n}\left(\mathbf{k}\right)\frac{\left\langle u_{\mathbf{k}}^{v}\left|\left[H,x\right]\right|u_{\mathbf{k}}^{c}\right\rangle }{E_{k}^{v}-E_{k}^{c}}k\,dk\,d\theta\right|^{2}}{E_{n}-\left(\hbar\omega+i\Gamma_n\right)}+\nonumber\\
	&\qquad+\left(\omega\rightarrow-\omega\right)^{*}.
\end{align}

The optical selection rules are directly obtained from the phase factors of the single particle states in Eq. (\ref{eq:eigenvectors_simple}) when the commutator $ \left\langle u_{\mathbf{k}}^{v}\left|\left[H,\mathbf{r}_{\alpha}\right]\right|u_{\mathbf{k}}^{c}\right\rangle $ is explicitly expanded. Recalling Eq. (\ref{eq:momentum_matrix}), as well as the discussion regarding eigenvector orthogonality that followed, the allowed transitions are associated with states with angular momentum $ \ell=0 $ ($ s $-series states) and $ \ell=\pm 2 $ ($ d $-series states). Explicitly, the commutator reads
\begin{equation}
	\left\langle u_{\mathbf{k}}^{v}\left|\left[H,x\right]\right|u_{\mathbf{k}}^{c}\right\rangle=\mathcal{A}\left(k\right)+\mathcal{B}\left(k\right)e^{-2i\theta},
\end{equation}
with $ \mathcal{A}\left(k\right)$ and $\mathcal{B}\left(k\right) $ the radial dependence of both the spinor components and the numerical parameters in the momentum matrix of Eq. (\ref{eq:momentum_matrix}). These two functions are then given by 
\begin{align}
	\mathcal{A}\left(k\right) &= -i v \eta\, a_{+}^{\dagger}\left(k\right)b_{-}\left(k\right)+\nonumber\\
	&\qquad+k\alpha\left[b_{+}^{\dagger}\left(k\right)b_{-}\left(k\right)-a_{+}^{\dagger}\left(k\right)a_{-}\left(k\right)\right], \nonumber\\
	\mathcal{B}\left(k\right) &= i v \eta\, b_{+}^{\dagger}\left(k\right)a_{-}\left(k\right)+\\
	&\qquad+k\alpha\left[b_{+}^{\dagger}\left(k\right)b_{-}\left(k\right)-a_{+}^{\dagger}\left(k\right)a_{-}\left(k\right)\right]\nonumber. 
\end{align}
To compare the oscillator strength of the two possible types of transitions, we compute the oscillator strength for $ d $-series transitions as
\begin{equation}
	\left|\boldsymbol{\Omega}_{n,d;x}\right|^2=\left|\int\frac{f_{n;d}\left(k\right)e^{2i\theta}\mathcal{B}\left(k\right)e^{-2i\theta} }{E_{k}^{v}-E_{k}^{c}}k\,dk\,d\theta\right|^{2},
\end{equation}
where $f_{n;d}\left(k\right)$ is the excitonic radial wave function for $ d $-series states. On the other hand, the oscillator strength for $ s $-series transitions is given by 
\begin{equation}
	\left|\boldsymbol{\Omega}_{n,s;x}\right|^2=\left|\int\frac{f_{n;s}\left(k\right)\mathcal{A}\left(k\right) }{E_{k}^{v}-E_{k}^{c}}k\,dk\,d\theta\right|^{2},
\end{equation}
where $f_{n;s}\left(k\right)$ is the excitonic radial wave function for $ s $-series states. The first resonance for the $ \ell=2 $ angular momentum series occurs at around $ 156\,\mathrm{meV} $, and we obtain an oscillator strength around $ 2 $ orders of magnitude smaller than that of the $ s $-series transitions closest to it. 

\begin{figure*}
	\centering{}\includegraphics{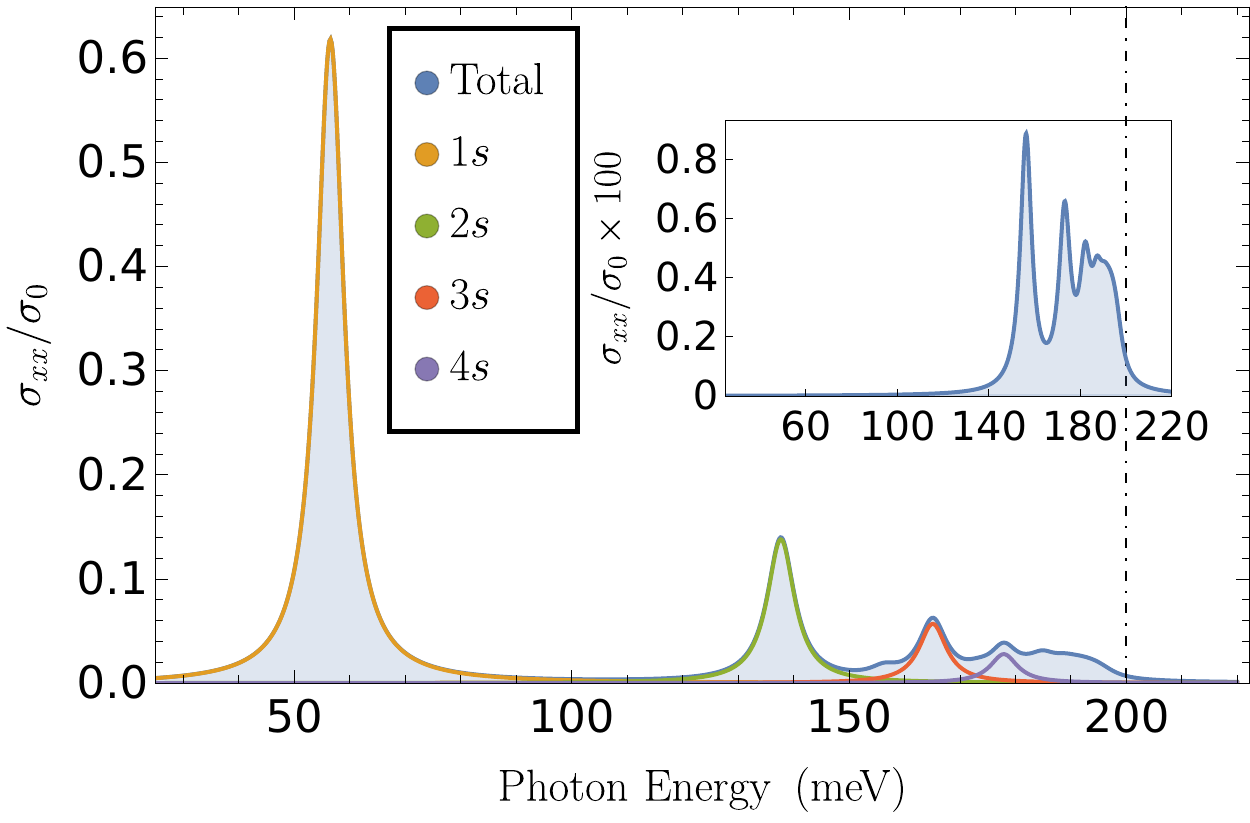}
	
	\caption{Real part of the excitonic $xx$-conductivity for a material described by a tilted Dirac Hamiltonian encapsulated in quartz with broadening parameter $\Gamma=3\,\mathrm{meV}$, and a $N=450$ point Gauss-Legendre quadrature. First ten states of each excitonic series were considered for the total conductivity. Vertical dashed lines represent the bandgap of the system. The conductivity is given in units of the conductivity of monolayer graphene $ \sigma_0 = e^2 / 4\hbar $. In the inset, we plot the contribution from only $ d $-series states scaled by two orders of magnitude to improve readability and comparison of the relative intensity. The vertical dashed line representing the bandgap has been aligned by the same value in both the main plot and the inset. \label{fig:conductivity_plot_inset}}
	
\end{figure*}

In Fig. \ref{fig:conductivity_plot_inset}, we plot the real part of the excitonic $xx$-conductivity with a broadening parameter of $\Gamma=3\,\mathrm{meV}$. In its inset, the contribution from $ d $-series states is also plotted, magnified by a factor of $ 100 $ as to improve comparison of the oscillator strengths of both types of transitions. 

\section{Conclusions}

We studied the properties of excitons in Janus TMD monolayers 
modeled by a tilted massive Dirac Hamiltonian. We have shown that, as the tilt 
parameter increases, the band gap is continuously suppressed, whereas the
maximum of the valence band and the minimum of the conduction band shift
in opposite directions along the tilting axis, making the gap indirect.
Notably, the exciton binding energies remain unchanged as the tilting
is enhanced. This means that the (indirect) gap can be made smaller than the exciton
binding, a situation that has been predicted to lead to an excitonic instability,
and possibly the formation of an excitonic insulator phase. 

Finally, we also considered the excitonic linear conductivity, discussing the optical 
selection rules for the system. With the model Hamiltonian considered, only states with 
angular momentum $ \ell=0 $ or $ \left|\ell\right|=2 $ can be excited, with the resonances 
associated with $ \ell=0 $ transitions more than two orders of magnitude greater than 
those associated with $ \ell=\pm 2 $ transitions. As expected from the solutions of the 
Bethe-Salpeter equation, the excitonic linear conductivity was also fully independent 
of the tilt parameter, even when comparing light polarized in either the parallel or 
the perpendicular direction of the tilt axis.
Importantly, the excitonic contribution to the conductivity is insensitive to
the value of the tilting parameter $t$. As the system is pushed towards the excitonic 
instability by tuning $t$, one can expect $\sigma_{xx}(\omega)$ to remain unchanged
until the system reaches the instability point, where $\sigma_{xx}(\omega)$ is
expected to change abruptly, marking the onset of the phase change.

\section*{Acknowledgements}
We acknowledge fruitful discussions with Joaqu\'in Fern\'andez-Rossier and Gon\c calo Catarina.
M. F. C. M. Q. acknowledges the International Nanotechnology Laboratory (INL) and the Portuguese Foundation for Science and Technology (FCT) for the Quantum Portugal Initiative (QPI) grant SFRH/BD/151114/2021. 
N. M. R. P. acknowledges support by the Portuguese Foundation for Science and Technology (FCT) in the framework of the Strategic Funding UIDB/04650/2020, COMPETE 2020, PORTUGAL 2020, FEDER, and  FCT through projects POCI-01-0145-FEDER-028114, POCI-01-0145-FEDER-02888 and PTDC/NANOPT/ 29265/2017, PTDC/FIS-MAC/2045/2021, EXPL/FIS-MAC/0953/ 2021, and from the European Commission through the project Graphene Driven Revolutions in ICT and Beyond (Ref. No. 881603, CORE 3).

\section*{References}
\bibliographystyle{unsrt}
\bibliography{main_v7.bib}

\end{document}